\documentclass[aps,pre,showpacs,10pt]{revtex4}
\usepackage[dvips]{epsfig}
\usepackage[dvips]{color}
\usepackage{graphicx}
\usepackage{float}
\usepackage{subfigure}

\begin{document}

\title{Modified Jeans Instability Criteria for Magnetized Systems}
\author{J. Lundin, M. Marklund, and G. Brodin}
\affiliation{Department of Physics, Ume\aa\ University, SE-901 87 Ume\aa,
Sweden}
\begin{abstract}
The Jeans instability is analyzed for dense magnetohydrodynamic plasmas with
intrinsic magnetization, the latter due to collective electron spin effects.
Furthermore, effects of electron tunneling as well as the Fermi pressure are
included. It is found that the intrinsic magnetization of the plasma will
enhance the Jeans instability, and can significantly modify the structure of
the instability spectra. Implications and limitations of our results are
discussed, as well as possible generalizations.
\end{abstract}
\maketitle

\section{Introduction}

Whenever the internal pressure of a gas or plasma cloud is too weak to
balance the self-gravitational force of a mass density perturbation, there
will be a collapse. The study of such mechanisms was pioneered by Jeans, and
it is well established that the Jeans instability plays a crucial role in
the mechanism responsible for the formation of structures in the universe 
\cite{Jeans1929}.

For lengthscales comparable to the de Broglie wavelength
of the charge carrier, tunneling effects become important and give rise to dispersion. This is particularly important for low temperature and/or high density plasmas. A characteristic
condition for this effect to be of significance is when the de Broglie
wavelength is comparable or larger than the mean separation of particles.
The plasma then behaves as a Fermi gas, obeying Fermi-Dirac statistics
rather than Boltzmann-Maxwell used in classical plasmas.
These effects can be captured within the magnetohydrodynamic (MHD) model through a modified equation
of state for the pressure \cite{Manfredi2005,PLA16}, and by including the Bohm potential \cite
{Manfredi2005,PLA7,Haas2005,PLA9,PLA10} describing quantum forces due to non-locality
effects. Quantum plasmas was first studied by Pines in the 1960's \cite{Pines1961,Pines1999}, and many studies has appeared since then \cite{PLA3}, e.g., kinetic models of the quantum electrodynamical properties of nonthermal plasmas \cite{PLA4} and covariant Wigner function descriptions of relativistic quantum plasmas \cite{PLA5}. The study of quantum plasmas in recent years have been motivated by e.g.\ possible applications to nanoscale technology \cite{pop4}, new developments in microelectronics \cite{Markovich}, the discovery  of ultracold plasmas \cite{pop6,Killian, Li} and the experimental demonstration \cite{pop7} of collective modes in ultra cold plasmas, new laser plasma/solid-matter interaction regimes offered by the next generation of high intensity light sources \cite{Kremp1999,Andreev,pop9,pop10}, as well as developments in laser fusion \cite{pop8}. Quantum effects are also believed to be of importance in the interiors of compact astrophysical objects \cite{Jung2001,Opher2001, Chabrier2002} such as white dwarfs, neutron stars, magnetars and supernovas, where the density can reach ten orders of magnitude that of ordinary solids. Furthermore, in some astrophysical environments, such as in the vicinity of pulsars \cite
{Beskin1993,Asseo2003} and magnetars \cite{Kouveliotou1998}, strong external
magnetic fields may be present. In such dense and/or strong magnetic field environments a quantum description
of the plasma which incorporates the spin of the particle is needed. A great
deal of interest has been directed toward finding such quantum plasma
descriptions \cite{Baring2005,Harding2006,Brodin2007,PLA14,PLA13,PLA17,PLA18}. In the context of the Jeans instability, quantum effects have been studied in Ref.\ \cite{Shukla2006}.

The objective of this paper is to investigate the Jeans instability in a
magnetized self-gravitating spin--$1/2$ plasma. Starting with the
MHD equations, Poisson's equation for the gravitational
potential, and an expression for the magnetization of the plasma due to the
electron spin, we derive a dispersion relation for plasma modes with
arbitrary directions of propagation. The overall stability of the system is
investigated, and in particular we study the stabilizing/destabilizing
effect of the Bohm potential and the electron spin. Finally, we discuss the implications and limitations of
our model.

\section{Governing equations}

The governing equations for self-gravitating spin--$1/2$ ideal MHD plasmas \cite{Haas2005,Brodin2007,PLA35,Marklund2007}
are the mass density conservation law 
\begin{equation}
\partial _{t}\rho +\nabla \cdot (\rho \mathbf{v})=0,
\end{equation}
the momentum equation 
\begin{equation}
\rho (\partial _{t}+\mathbf{v}\cdot \nabla )\mathbf{v}=-\rho \nabla \phi
-\nabla p-\nabla \left( \frac{B^{2}}{2\mu _{0}}-\mathbf{M}\cdot \mathbf{B}%
\right) +\mathbf{B}\cdot \nabla \left( \frac{\mathbf{B}}{\mu _{0}}-\mathbf{M}%
\right) +\frac{\hbar ^{2}\rho }{2m_{e}m_{i}}\nabla \left( \frac{\nabla ^{2}%
\sqrt{\rho }}{\sqrt{\rho }}\right) ,  \label{eq:momentum}
\end{equation}
the idealized Ohm's law 
\begin{equation}
\partial _{t}\mathbf{B}=\nabla \times (\mathbf{v}\times \mathbf{B}),
\end{equation}
and Poisson's equation for the gravitational potential 
\begin{equation}
\nabla ^{2}\phi =4\pi G\rho .
\end{equation}
Here $\rho $ is the mass density, $\mathbf{v}$ is the fluid velocity, $%
\mathbf{B}$ is the magnetic field, $\mathbf{M}$ is the magnetization due to
the electron spin, $\phi $ is the gravitational potential determined by
Poisson's equation, $\hbar $ is Planck's constant divided by $2\pi $, $%
m_{e,i}$ is the electron and ion mass respectively, and $G$ is the
gravitational constant. The momentum equation, eq.\ (\ref{eq:momentum}), has
been modified to include non-locality effects, such as tunneling, described
by the Bohm potential. These effects are important for dense and/or low temperature plasma systems
and for short wavelengths.

This system of equations is closed with an equation of state for the
pressure and an expression for the magnetization. The equation of state for
the pressure is written as, $\nabla P=c_{s}^{2}\nabla \rho $, where $%
c_{s}=(dp/d\rho _{0})^{1/2}$ is the sound speed. The ion-acoustic velocity
can be written as 
\begin{equation}
c_{s}^{2}=v_{ti}^{2}+\frac{m_{i}}{m_{e}}\left( v_{te}^{2}+\frac{3}{5}%
v_{Fe}^{2}\right) ,
\end{equation}
where $v_{ti}$ and $v_{te}$ are the ion and electron thermal velocities and $%
v_{Fe}=\hbar (3\pi ^{2}n_{e})^{1/3}/m_{e}$ is the electron Fermi velocity.
For temperatures well below the Fermi temperature, the thermal velocity is
given by $v_{ts}=C(k_{B}T/\hbar n_{s}^{1/3})$, where $C$ is a dimensionless
constant of order unity, whereas for temperatures much higher than the Fermi
velocity we have $v_{ts}=(k_{B}T/m_{s})^{1/2}$. Here $n_s$ is the number density of particle species $s$, $k_{B}$ is Boltzmann's
constant and $T$ is the temperature.

Multifluid equations including the effect of the magnetization due to the
electron spin has been derived in Ref.\ \cite{Marklund2007} and further
developed for MHD regimes in Ref.\ \cite{Brodin2007}. It is then showed that
for dynamics on a time scale much slower than the spin precession frequency,
the magnetization is given by 
\begin{equation}
\mathbf{M}=\frac{\mu _{B}\rho }{m_{i}}\tanh \left( \frac{\mu _{B}B}{k_{B}T}%
\right) \hat{\mathbf{B}},
\end{equation}
where, $\hat{\textbf{B}}=\textbf{B}/B$, and $B=\left|\textbf{B}\right|$. Here it has been assumed that the spin orientation has reached the thermodynamic
equilibrium state in response to the magnetic field, which accounts for the $%
\tanh (\mu _{B}B/k_{B}T)$-factor, where $\mu_B=e\hbar/2m_e$ is the Bohr magneton. Furthermore, on a time scale shorter than
the spin relaxation time scale, the individual electron spins are conserved,
and thus we can take $\tanh (\mu _{B}B/k_{B}T)$ as constant for an initially
homogeneous plasma \cite{BMM2008}.

Assuming perturbations on a homogeneous background (equilibrium values
denoted by index 0) we write $\rho =\rho _{0}+\delta \rho $, $\phi =\phi
_{0}+\delta \phi $, $\mathbf{v}=\delta \mathbf{v}$, $\mathbf{B}=B_{0}\hat{%
\mathbf{z}}+\delta \mathbf{B}$, and $\mathbf{M}=M_{0}\hat{\mathbf{z}}+\delta 
\mathbf{M}$ and linearize our governing equations. Making a harmonic
decomposition, we obtain 
\begin{equation}
\omega \delta \rho -\rho _{0}\mathbf{k}\cdot \delta \mathbf{v}=0,
\label{eq:linj-continuity}
\end{equation}
\begin{equation}
-i\omega \rho _{0}\delta \mathbf{v}=-i\mathbf{k}\left( \rho _{0}\delta \phi
+c_{s}^{2}\delta \rho +\frac{B_{0}\delta B_{z}}{\mu _{0}}-M_{0}\delta
B_{z}-B_{0}\delta M_{z}\right) +ik_{z}B_{0}\left( \frac{\delta \mathbf{B}}{%
\mu _{0}}-\delta \mathbf{M}\right) -\frac{i\hbar ^{2}k^{2}\mathbf{k}}{%
4m_{e}m_{i}}\delta \rho ,  \label{eq:linj-fluid-force}
\end{equation}
\begin{equation}
-\omega \delta \mathbf{B}=B_{0}\mathbf{k}\times (\delta \mathbf{v}\times 
\hat{\mathbf{z}}),  \label{eq:linj-mhd}
\end{equation}
\begin{equation}
-k^{2}\delta \phi =4\pi G\delta \rho ,  \label{eq:linj-grav}
\end{equation}
and 
\begin{equation}
\delta \mathbf{M}=M_{0}\left( \frac{\delta \mathbf{B}_{\perp }}{B_{0}}+\frac{%
\delta \rho }{\rho _{0}}\hat{\mathbf{z}}\right) ,  \label{eq:linj-magn}
\end{equation}
where $\delta \mathbf{B}_{\perp }=(\delta B_{x},\delta B_{y},0)$, and $%
M_{0}=(B_{0}/\mu _{0})\chi /(1+\chi )$. Here 
\begin{equation}
\chi =\frac{\frac{B_{0}\mu _{B}}{m_{i}C_{A}^{2}}\tanh \left( \frac{\mu
_{B}B_{0}}{k_{B}T}\right) }{1-\frac{B_{0}\mu _{B}}{m_{i}C_{A}^{2}}\tanh
\left( \frac{\mu _{B}B_{0}}{k_{B}T}\right) }
\end{equation}
is the magnetic susceptibility and $C_{A}=\left( B_{0}^{2}/\mu _{0}\rho
_{0}\right) ^{1/2}$ is the Alfv\'{e}n velocity.

Eq.\ (\ref{eq:linj-mhd}) already gives us an expression for $\delta \mathbf{B%
}$ in terms of $\delta \mathbf{v}$ and we can solve eq.\ (\ref
{eq:linj-continuity}) for $\delta \rho $, and then solve eq.\ (\ref
{eq:linj-grav}) for $\delta \phi $. Using these results in eq.\ (\ref
{eq:linj-fluid-force}) we find an equation for $\delta \mathbf{v}$. The
parallel and perpendicular component of $\delta \mathbf{v}$ are given by 
\begin{equation}
-\omega ^{2}\delta v_{z}=-\left( -\frac{4\pi G\rho _{0}}{k^{2}}+c_{s}^{2}+%
\frac{\hbar ^{2}k^{2}}{4m_{e}m_{i}}\right) k_{z}\mathbf{k}\cdot \delta 
\mathbf{v}+\frac{\chi }{1+\chi }C_{A}^{2}k_{z}\mathbf{k}\cdot \delta \mathbf{%
v}_{\perp } \label{eq:vz}
\end{equation}
and 
\begin{equation}
-\omega ^{2}\delta \mathbf{v}_{\perp }=-\mathbf{k}_{\perp }\left( -\frac{%
4\pi G\rho _{0}}{k^{2}}\mathbf{k}+c_{s}^{2}\mathbf{k}+\frac{1}{1+\chi }%
C_{A}^{2}\mathbf{k}_{\perp }-\frac{\chi }{1+\chi }C_{A}^{2}\mathbf{k}+\frac{%
\hbar ^{2}k^{2}}{4m_{e}m_{i}}\mathbf{k}\right) \cdot \delta \mathbf{v}-\frac{%
1}{1+\chi }C_{A}^{2}k_{z}^{2}\delta \mathbf{v}_{\perp } \label{eq:vorto}.
\end{equation}
Eqs.\ (\ref{eq:vz}) and (\ref{eq:vorto}) can be written in matrix form as $%
D^{ab}v_{b}=0$, and the dispersion relation is obtained from $\Vert
D^{ab}\Vert =0$, 
\begin{eqnarray}
\left( \omega ^{2}-\frac{1}{1+\chi }C_{A}^{2}k_{z}^{2}\right) \left\{ \left[
\omega ^{2}-\frac{1}{1+\chi }C_{A}^{2}k^{2}-\left( V^{2}-\frac{\chi }{1+\chi 
}C_{A}^{2}\right) k_{\perp }^{2}\right] \left( \omega ^{2}-V^{2}k_{z}^{2}%
\right) -k_{\perp }^{2}k_{z}^{2}\left( V^{2}-\frac{\chi }{1+\chi }%
C_{A}^{2}\right) ^{2}\right\} &=&0.  \nonumber  \label{eq:disprel} \\
&&
\end{eqnarray}
Here we have used the notation, 
\begin{equation}
V^{2}\equiv -\frac{4\pi G\rho _{0}}{k^{2}}+c_{s}^{2}+\frac{\hbar ^{2}k^{2}}{%
4m_{e}m_{i}},  \label{Eq: V-definition}
\end{equation}
The first factor of the dispersion relation describes the shear Alfv\'{e}n
wave, which is always stable. Focusing on the stability properties, this mode
is not considered below. The second factor describes the fast and the
slow magnetosonic modes modified by a gravitational potential, non-locality
effects due to the Bohm potential, and the spin magnetization. Our
dispersion relation agrees with that found in Refs.\ \cite{Brodin2007} and 
\cite{PLA17} in the appropriate limits if we correct for a sign error
in those papers.

In order to gain some understanding of the stability properties, we first
study the special cases of parallel and orthogonal propagation to the background magnetic field. 
Initially we discard the contribution $\hbar ^{2}k^{2}/4m_{e}m_{i}$ in (\ref{Eq:
V-definition}) due to the Bohm potential, which only becomes important for
short wavelengths. For parallel propagation, the instability condition then
simply becomes the ordinary Jeans instability condition, 
\begin{equation}
\frac{4\pi G\rho _{0}}{k^{2}}>c_{s}^{2}.  \label{Eq:Jeans-orginal}
\end{equation}
Formally, this means that our system is always unstable for sufficiently
long wavelengths. However, we may interpret the condition (\ref
{Eq:Jeans-orginal}) such as to give a required size of the system for an
instability of this type to be possible. In particular, for the above
calculation to be applicable, the system must be approximately homogeneous
over length scales of the order of the Jeans length, $\lambda _{J}=2\pi
/k_{J}$, where $k_{J}=(4\pi G\rho _{0}/c_{s}^{2})^{1/2}$. Next we consider
orthogonal propagation, the instability condition becomes, 
\begin{equation}
V^{2}+\frac{1-\chi }{1+\chi }C_{A}^{2}<0.  \label{Eq:Magn-inst}
\end{equation}
In this case the system may be unstable even if the gravitational effect is
omitted, provided the condition 
\begin{equation}
\frac{\chi }{1+\chi }>\frac{1}{2}\left( 1+\frac{c_{s}^{2}}{C_{A}^{2}}\right) \label{eq:cond3}
\end{equation}
is fulfilled. The condition (\ref{eq:cond3}) then suggests that the plasma is unstable for all length scales. In
particular even in the short wavelength regime, where our calculation is of special interest 
if we assume that the plasma system of study has a limited
extension. For short wavelengths, on the other hand, we expect the Bohm
potential to be of importance. Noting from (\ref{Eq: V-definition}) and (\ref
{Eq:Magn-inst}) that the Bohm potential acts stabilizing for sufficiently
short wavelengths, this term is therefore kept below.

Next we normalize the parameters according to 
\[
\bar{\omega}=\frac{\omega }{k_{J}c_{s}},\qquad \bar{k}=\frac{k}{k_{J}}%
,\qquad \alpha =\frac{\chi }{1+\chi },\qquad \bar{C}_{A}=C_{A}/c_{s},\qquad 
\bar{Q}=\frac{\hbar ^{2}k_{J}^{2}}{4m_{e}m_{i}c_{s}^{2}}, 
\]
such that the dispersion relation can be written as 
\begin{eqnarray}
\bar{\omega}^{4}-\bar{\omega}^{2}\left[ \left( -\frac{1}{\bar{k}^{2}}+1+%
\bar{Q}\bar{k}^{2}+\bar{C}_{A}^{2}\left( 1-\alpha \right) \right) \bar{k}%
^{2}-\alpha \bar{C}_{A}^{2}\bar{k}_{\perp }^{2}\right] &+&  \nonumber \\
\bar{k}_{z}^{2}\bar{C}_{A}^{2}\left[ \left( -\frac{1}{\bar{k}^{2}}+1+\bar{Q}%
\bar{k}^{2}\right) \left( \bar{k}^{2}-\alpha \bar{k}_{z}^{2}\right) -\alpha
^{2}\bar{C}_{A}^{2}\bar{k}_{\perp }^{2}\right] &=&0.
\end{eqnarray}
The growth rate $\gamma =\text{Im }\bar{\omega}$ as a function of $\bar{k}%
_{z}$ and $\bar{k}_{\perp }$ has been illustrated in Fig.\ 1 for $(a)$ the
classical Jeans instability and $(b)$ the Jeans instability modified by the
presence of an external magnetic field. The external magnetic field will
decrease the instability in the direction orthogonal to the magnetic field.
This will generate an oblate spheroidal collapse. This is expected since the
magnetic forces are anisotropic so the magnetic pressure primarily work in
directions perpendicular to $\mathbf{B}$. While the oblate shape of the
plasma cloud is an effect that occur in the later (nonlinear) stage of the
collapse, we note that the stabilizing influence of the magnetic field can
be seen already in the initial linear stage of the instability. In
particular the dependence of the growth rate on the wavenumber shows that
perturbations perpendicular to the magnetic field is less unstable, as is
seen by comparing Fig.\ $(1a)$ and Fig.\ $(1b)$. Furthermore, we note that the
electron non-locality effects described by he Bohm potential has an over all
stabilizing effect of the system which is illustrated in Fig.\ $(1c)$.

Effects due to the electron spin has a destabilizing effect on the system,
in particular in directions almost, but not quite, perpendicular to the
magnetic field. This instability stems from the magnetic attraction of
individual spins (i.e.\ magnetic dipole moments) and has been investigated in
Ref.\ \cite{PLA17}. The effect of electron spin on the Jeans
instability is demonstrated in Fig.\ $(1d)$. In this parameter regime the
spin magnetization is significant, but the plasma would still be stable if
gravitational effects were disregarded. This is in contrast to Fig.\ $(1e)$ 
in which the spin contribution alone is sufficient for collapse. Here,
stability is regained for short wavelengths through the influence of the
Bohm potential. For comparison, the spin instability alone, i.e.\ without the
gravitational Jeans contribution, has been illustrated in Fig.\ $(1f)$ with
the same parameter values as in Fig.\ $(1e)$.

The maximum value of $\left| \mathbf{k}\right| =k_{\text{max}}$ for which we
have instability occur at an angle to $\mathbf{k}_{\perp }$, as can be seen
in e.g.\ Fig.\ $(1d)$. Our investigation is of particular interest when $k_{%
\text{max}}>k_{J}$. Especially when considering a system of finite size,
with inhomogeneity scale length shorter than the Jeans length, i.e.\ for $%
\lambda _{J}\left( \nabla n_{0}/n_{0}\right) $ larger than unity. In this
regime there will typically be no collapse due to the Jeans instability
alone, but as shown from the above analysis a collapse for short wavelengths
(which is well described by the homogeneous background model) can still
occur when gravitational and magnetization effects are combined.

\section{Summary and Discussion}

In the present study the effect of magnetization due to the electron spin
has been combined with Newtonian gravity, in order to describe the stability
properties of an initially homogeneous plasma. We note that a more complete treatment of the Jeans instability would require taking cosmological effects into account \cite{Peebles}. As expected, the
gravitational effects tend to dominate on the large scales, i.e.\ for long
wavelengths. For systems of a limited size, background inhomogeneities not
included in our model can stabilize the longest wavelengths, in which case
more significance is given to the short wavelength properties of the
dispersion relation. Furthermore, for a system of moderate temperature and
high density, it is seen that the magnetization of the plasma can contribute
significantly to the instability in the short wavelength regime. For short
wavelengths, we also note that the Bohm potential can be of importance,
providing a stabilizing influence. A case where the present study is of
special interest is when the long wavelength perturbations are stabilized by
inhomogeneities, and the magnetization of the plasma is significant but not
sufficient to cause an instability by itself. For such a plasma, the
combined effects of gravitation and magnetization is needed to fulfill the
instability condition. 

In the present model we have not included the effect of the Hall current 
\cite{Brodin-Stenflo1990}, or other finite Larmor radius
effects \cite{Brodin-Stenflo1988} that is often introduced to improve the
ideal MHD equations. We note that such modifications introduce dispersive
properties of the waves, similar to the effects of the Bohm potential.
However, as far as the stability properties is concerned, the moderate
temperature high density regime is of most interest, since it is in this
regime that the spin effects can give a significant contribution to the
instability, as seen above. Furthermore, for such parameters the effects of
the Bohm potential is more important than those of the Hall current. 

Naturally there are several processes that can be expected in a physical
collapse that are not captured within the low-frequency MHD limit. For
instance, electrostatic repulsion, that could have a stabilizing effect on
the system, is not included due to the assumption of quasineutrality. 
Furthermore, since our ideal MHD equations have an infinite conductivity,
magnetic field lines will be frozen-in during collapse. In reality, there
will be particle collisions within the plasma and thus the conductivity will
always remain finite. The finite conductivity will prevent strict
flux-freezing, and as a result the magnetic field will not grow as fast as
MHD predicts. This will also reduce the effects of spin magnetization
somewhat.

As noted \ above, magnetization effects as well as electron tunneling and
exclusion principle effects, are particularly important in low temperature
high density systems. In particular only electrons with energies larger than
the Fermi energy will be free to contribute to the magnetization process.
Thus, the magnetization will need to contain an effective density of
electrons, reduced by the number of electron below the Fermi surface. As
such, a different effective scaling of the magnetization as a function of
density will be obtained. Moreover, correlation effects in dense plasmas is
of importance, as well as relativistic effects. These are examples of
possible modifications that could be introduced in modeling of such quantum
plasmas in the future.

The analysis here has been performed through a linearization of the
perturbed parameters. More sophisticated methods exists which can give
qualitative descriptions of the non-linear dynamics of a collapse. The use
of integral virial theorems, for instance, has proven to often give a more
reliable global view of the interaction of magnetic and gravitational fields
in the non-linear domain \cite{Mestel}. Still, for a more complete
understanding of a physically realistic collapse, the use of advanced
computer simulations is often an necessity. 


\begin{figure}[]
\begin{minipage}{0.4\linewidth}
\includegraphics[width=1\textwidth]{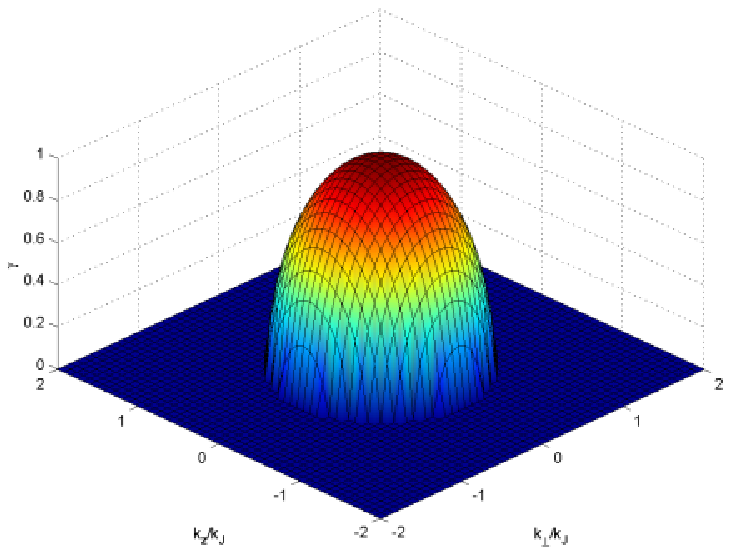}
\begin{footnotesize}
\begin{center}
\textbf{(a)}
\end{center}
\end{footnotesize}
\includegraphics[width=1\textwidth]{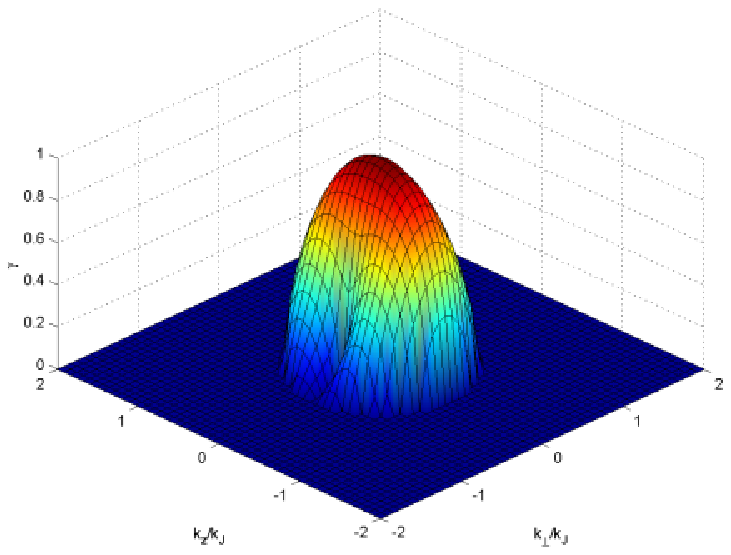}
\begin{footnotesize}
\begin{center}
\textbf{(c)}
\end{center}
\end{footnotesize}
\includegraphics[width=1\textwidth]{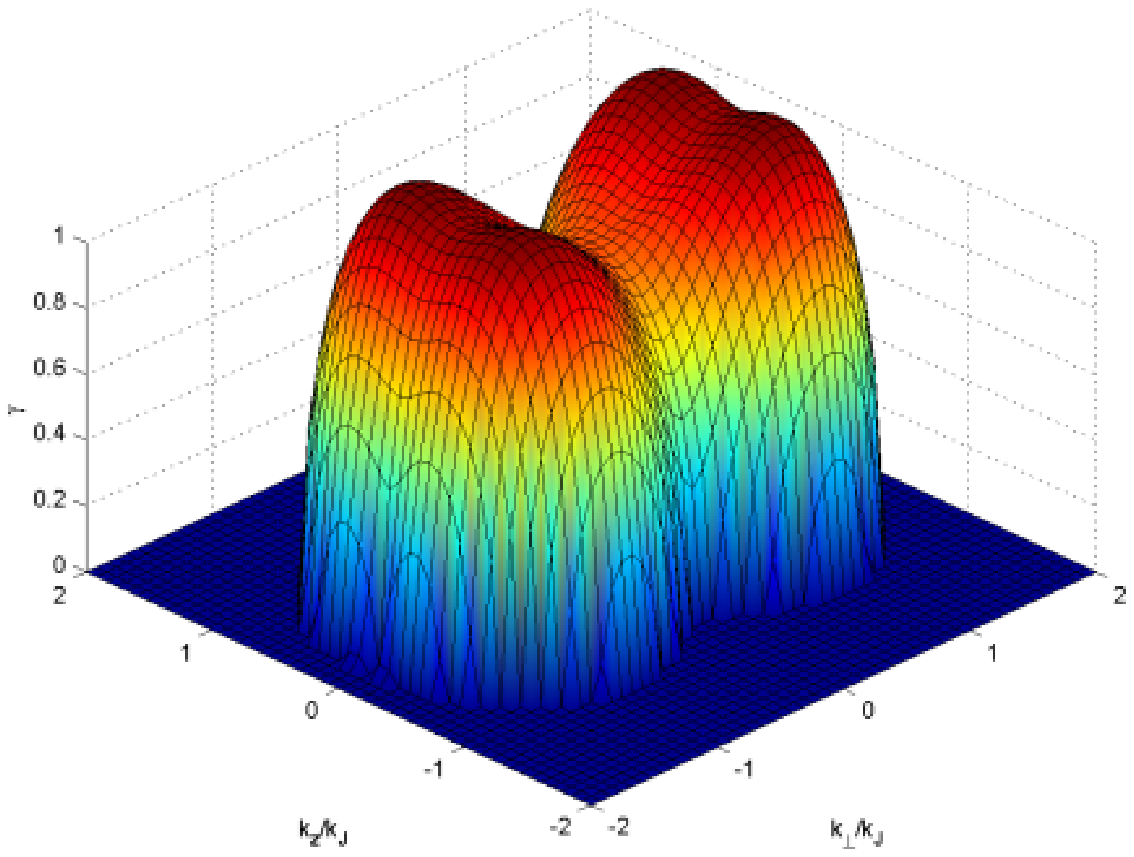}
\begin{footnotesize}
\begin{center}
\textbf{(e)}
\end{center}
\end{footnotesize}
\end{minipage}
\begin{minipage}{0.4\linewidth}
\includegraphics[width=1\textwidth]{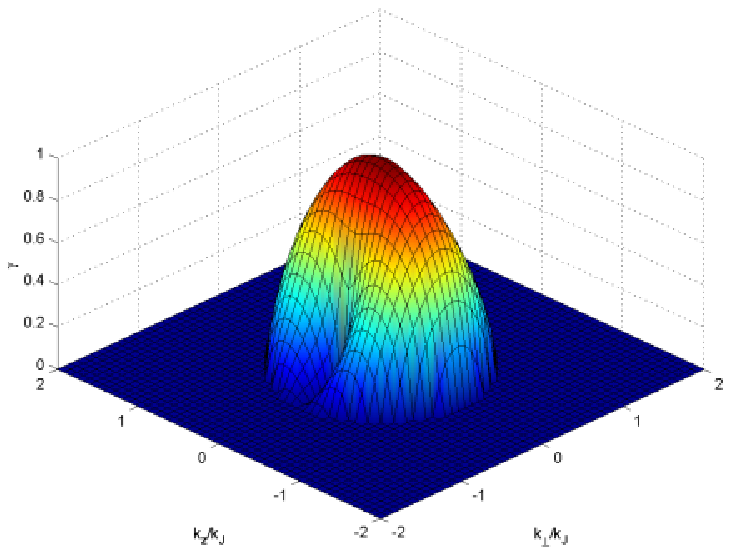}
\begin{footnotesize}
\begin{center}
\textbf{(b)}
\end{center}
\end{footnotesize}
\includegraphics[width=1\textwidth]{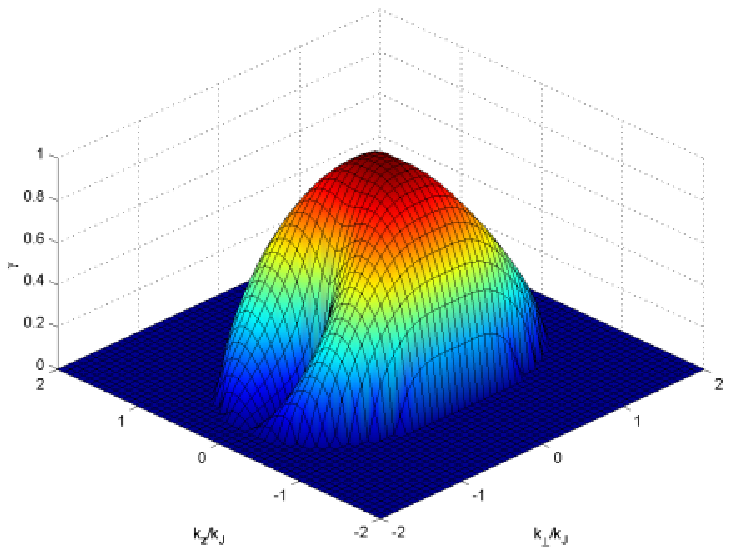}
\begin{footnotesize}
\begin{center}
\textbf{(d)}
\end{center}
\end{footnotesize}
\includegraphics[width=1\textwidth]{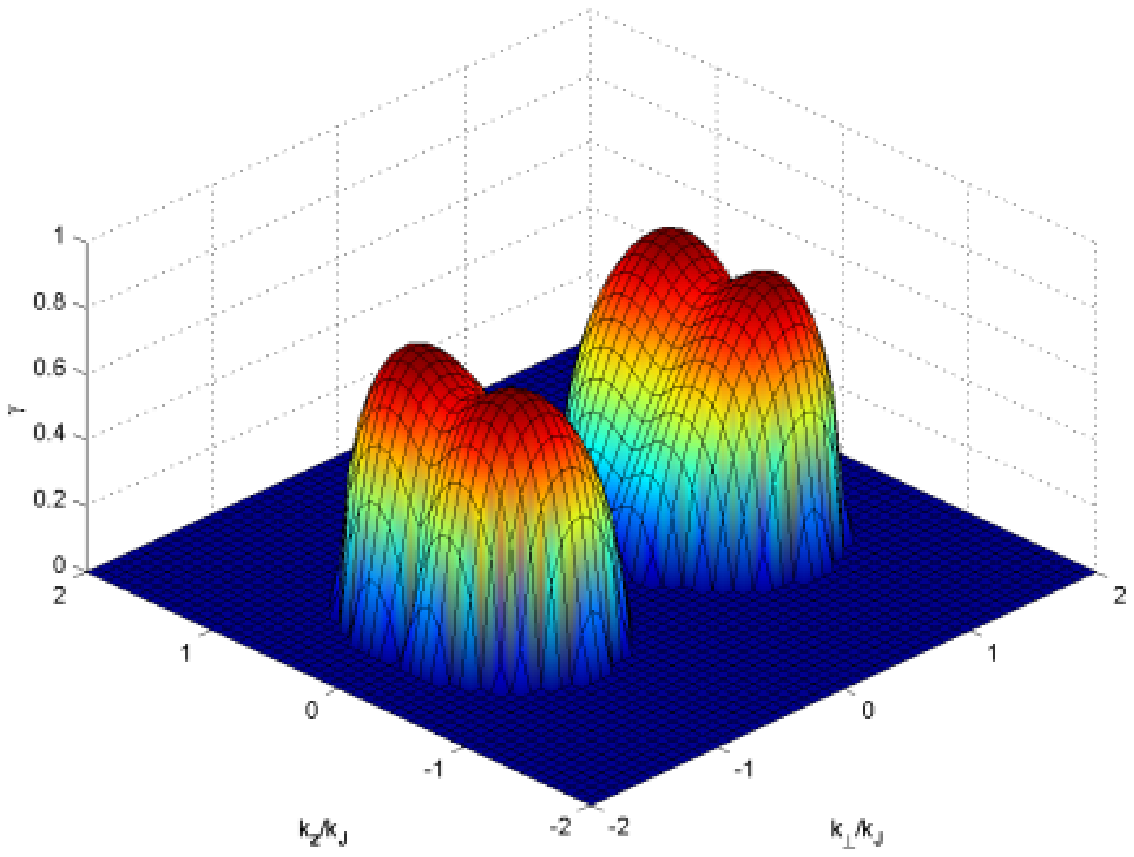}
\begin{footnotesize}
\begin{center}
\textbf{(f)}
\end{center}
\end{footnotesize}
\end{minipage}
\caption{The growthrate has been plotted for $(a)$ the classical Jeans
instability and (b) the Jeans instability modified by the presence of an
external magnetic field ($\bar C_A=2$ in Fig.\ $(b)-(f)$). In $(c)$ we have
added non-locality effects of the electrons described by the Bohm potential (%
$\bar Q=0.5$), and $(d)$ includes electron spin effects in the plasma
description ($\protect\alpha=0.4$). In $(e)$ both spin and non-locality
effects are included in the model in a regime ($\protect\alpha=0.75, \bar
Q=0.5$) where the plasma would be unstable even without graviational
effects, Fig.\ $(f)$.}
\label{fig:Compare1}
\end{figure}

\end{document}